# Plasmon Assisted Random Lasing of Perovskite Materials

Yagya Woli, Bryson Krause and Thang Hoang

*Department of Physics and Materials Science, The University of Memphis, Memphis, TN 38152*

Random lasing occurs as the result of coherent optical feedback from random scattering centers. Plasmonic nanostructures, such as silver or gold nanoparticles, efficiently scatter light due to the formation of hot spots and optical confinement at the nanoscale. In this work, using silver nanocubes as highly efficient light scattering centers, a broad band plasmon assisted lasing action of halide perovskite materials is demonstrated. By embedding silver nanocubes in $CH_3NH_3PbBr_3$ and $CH_3NH_3PbI_3$ solutions, narrow bandwidth lasing modes with a full width at half-maximum of approximately 1 nm are supported. It is observed that the lasing thresholds of perovskites are different for glass and gold substrates, and for different nanocube concentrations. Results of time-resolved measurements indicate a significant shortening in the decay time of the emission at above the lasing threshold, implying a stimulated emission process. The results of this work thus provide a pathway to generate coherent light sources from widely studied perovskite materials.

*Keywords: plasmonics, nanocubes, random lasing, nanolasers, perovskites*

Metal-halide perovskites have recently gained significant interest in the fields of photovoltaic and optoelectronic semiconductor materials. In a short period, the efficiency of halide perovskite solar cells has reached an impressive value of 25.2% [1]. Metal-halide perovskites are ideal for achieving optical gain in low-threshold lasing devices due to their simple structures, strong light absorption, tunable emission wavelengths, high carrier mobility, extended carrier diffusion length, easy preparation process, and excellent flexibility [2-5]. Lasing in perovskite materials was first reported in 1998 from layered $(C_6H_{13}NH_3)PbI_4$ polycrystalline films at 16 K [6]. Later, in 2014 optical gain at room temperature was achieved in solution processed methylammonium lead halide perovskite thin films [7]. Also in 2014, the first optically pumped perovskite microcavity laser was reported [8]. The lasing action of an embedded perovskite gain medium in periodic lattices of plasmonic nanoparticles has also been observed, with well-defined surface lattice plasmon resonance dependent on several material parameters such as dielectric constant, lattice structure, and nanoparticle size [9]. It is noteworthy that both ordered and randomly distributed plasmonic nanostructures have been shown to result in amplified spontaneous emission and coherent random lasing. The random lasing phenomenon is attributed to the multiple scattering processes, wherein the surface plasmon resonances of metallic nanoparticles act as scattering centers, forming random closed-loop cavities that effectively enhance the lasing process [10-14]. The rapid progress in perovskite photovoltaic and light-emitting devices has also led to a notable increase in interest in the use of hybrid perovskite materials for laser technology. So far, researchers have demonstrated optically pumped lasers employing a variety of optical feedback structures, including Fabry-Perot cavities [8,15], distributed feedback gratings [16], and whispering gallery modes [17].

Unlike traditional lasers, which depend on a resonator cavity with two precisely aligned mirrors to produce coherent optical feedback, random lasers function through a different feedback mechanism. In a random lasing system, feedback originates from multiple scattering processes, a common phenomenon in photonics. Indeed, random lasers have gained significant attention over the past two decades [18-20]. While periodic lattices of plasmonic nanoparticles provide well-determined surface plasmon resonance, which supports lasing action in embedded gain media, random lasing relies on localized surface plasmon resonances (LSPRs) [21,22] of metallic nanostructures. Indeed, LSPR has garnered significant attention due to its unique physical mechanism and potential applications [21-24]. In earlier findings, it has been shown that the lasing threshold was reduced, and its emission was significantly enhanced by the introduction of silver or gold nanoparticles [10,25-29]. In general metal nanoparticles play a significant role in the enhancement of lasing efficiency by two mechanisms [25]: the first aspect involves amplifying the localized electromagnetic field around the metal nanoparticle, while the second aspect involves enhancing scattering effects due to the significant scattering cross section of metal nanoparticles. Interestingly it has been demonstrated that a random distribution of plasmonic nanostructures can also lead to amplified spontaneous emission and coherent random lasing due to multiple scattering processes [11]. In this work the efficient random lasing process is experimentally demonstrated through the utilization of silver (Ag) nanocubes with the metal halide perovskite $MAPbBr_3$ and $MAPbI_3$ (MA = Methylammonium [$CH_6N+$]) crystals used as a gain medium. Lasing action and a significantly narrower spectral linewidth are observed in metal halide perovskite matrices containing silver nanocubes, in contrast to the broad spontaneous photoluminescence (PL) emission exhibited by perovskites without silver nanocubes. We further illustrate the impact of the substrate on the lasing threshold of metal halide perovskite coupled with Ag nanocubes. Our observations revealed a noteworthy reduction in the lasing threshold and a significant narrowing of the spectral lasing linewidth when employing a gold substrate, as opposed to a glass substrate.

Previous studies have indicated that Ag nanocubes have a broad absorption band as well as a broad scattering band [30-32]. In our study, 100 nm Ag nanocubes are embedded in a gain medium of either $MAPbBr_3$ or $MAPbI_3$. Under ambient conditions, crystalline $MAPbBr_3$ and $MAPbI_3$ perovskites exhibit PL emissions at wavelengths of approximately 550 nm and 780 nm, respectively. It is important to emphasize that the broad spectral band of Ag nanocubes aligns well with both these PL emission bands [30]. The presence of Ag nanocubes in the $MAPbBr_3$ or $MAPbI_3$ matrices is expected to enhance optical properties, elevate light-matter interactions, and create favorable conditions for lasing in perovskite materials. The specific influence will depend on factors such as the size and concentration of the Ag nanocubes, along with the overall design of the device.

In the experimental process, varying concentrations of Ag nanocubes are integrated into a solution of $MAPbBr_3$ or $MAPbI_3$. The $MAPbX_3$ (X = Br or I) solution is created by dissolving 0.25M of MAX and 0.25 $PbX_2$ in a ratio of 1.05:1, forming a mixture in 1 ml of DMF (Dimethylformamide) [33]. Mixtures $MAPbBr_3$ ($MAPbI_3$) were prepared with varying relative concentrations. Then various concentrations of Ag nanocubes are combined with 30 μl of $MAPbBr_3$ or $MAPbI_3$ solution individually. After thorough mixing through vortexing, a droplet of the $MAPbBr_3$ ($MAPbI_3$) and Ag nanocube mixture is deposited at the edge of a gold (100 nm thick) and/or glass substrate. The mixture of $MAPbBr_3$ ($MAPbI_3$) with Ag nanocubes is excited by a short-pulsed solid-state laser at 515 nm (Coherent Flare NX The laser has a pulse length shorter than 1.0 ns, a maximum pulse energy of 322 μJ, and operates at a repetition rate of 2 KHz). For efficient excitation and collection, the laser excitation beam was focused into a thin stripe, as depicted in Figure 1(a), by using a pair of cylindrical lenses. The excitation laser is directed from the top of the sample through a 10X objective lens (Mitutoyo), and the signal collection occurs from a side facet, resembling a setup analogous to a waveguide configuration [34]. Such an experimental configuration enables efficient excitation and facilitates collection into an optical fiber. The laser excitation spot size is approximately 165 μm² and all the laser excitation powers referred in this work were the averaged power measured before the objective lens. The lasing signal is filtered through a long-pass filter. Subsequently, the gathered lasing signal is directed into an optical fiber and examined by using a portable spectrometer (Ocean Optics USB2000+). In time-resolved single-photon counting measurements, the lasing signals are directed in free space to another spectrometer (Horiba iHR550) where they are guided through a second exit port and gathered by a fast-timing avalanche photodiode. A time-correlated single-photon counting module (PicoHarp 300 by Picoquant) with a time bin of 4 ps is employed to assess the number of photons concerning their arrival time at the photodiode. The computed lifetimes are ultimately derived from fitting the data after deconvolution with the instrument's response function.

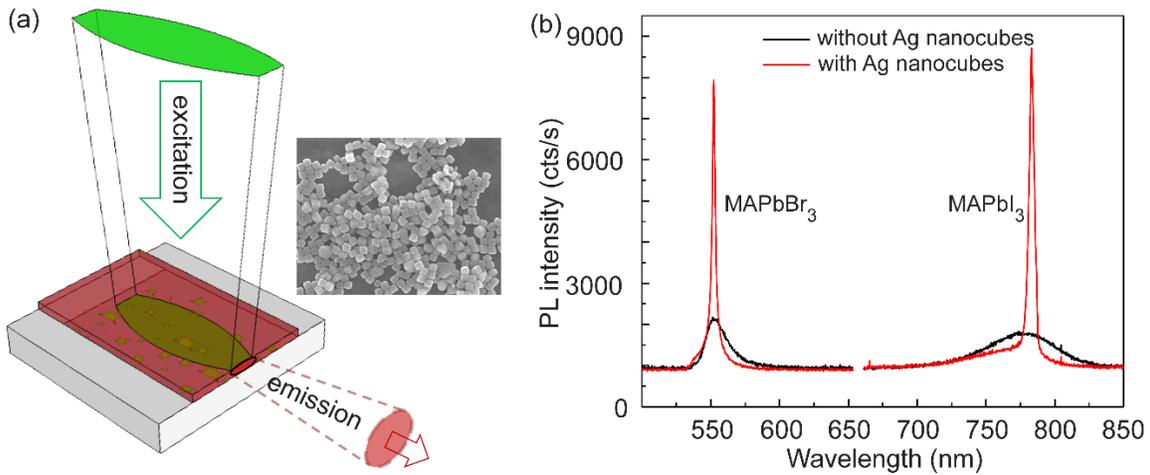

**Figure 1.** (a) Schematic of the experimental excitation and detection configuration. Inset: SEM image of Ag nanocubes used for lasing experiments. (b) PL emission spectra of MAPbBr$_3$ and MAPbI$_3$ with and without Ag nanocubes.

In this section the lasing action of metal halide perovskite ($MAPbBr_3$ or $MAPbI_3$) coupled with different concentrations of Ag nanocube on both gold and glass substrate is demonstrated. The plasmonic resonance exhibited by Ag nanocubes in $MAPbBr_3$ and/or $MAPbI_3$ matrix is a crucial factor contributing to the heightened absorption and scattering of light significantly influencing the overall optical characteristics of the material. Previous studies have demonstrated that Ag nanocubes have a strong interaction with light over a broad spectral range (400-1000 nm) from both absorption and scattering aspects.[30-32] The inset in Figure 1(a) shows an scanning electron microscope (SEM) image of Ag nanocubes on a Au substrate that was used in this work.

The lasing enhancement supported by the addition of Ag nanocubes is evident from the data shown in Figure 1(b) depicting the emission spectra of $MAPbBr_3$ and $MAPbI_3$ samples on glass substrates for with and without Ag nanocubes. At the same laser excitation power (~ 23 $W/cm^2$) emissions from both $MAPbBr_3$ and $MAPbI_3$ show no evidence of lasing action when Ag nanocubes are absent while clear narrow lasing peaks appear when Ag nanocubes are observed. Without Ag nanocubes, the $MAPbBr_3$ and $MAPbI_3$ samples show no evidence of lasing action, even when the laser excitation power is increased to 35 $W/cm^2$.

On a glass substrate at lower laser fluence, the emission from $MAPbBr_3$ coupled with Ag nanocubes is characterized by a broad spontaneous emission spectrum. However, as the laser fluence increases, a narrow-stimulated emission peak appears at 550 nm with a full width at half-maximum (FWHM) of about 1 nm, indicating lasing action as shown in Figure 2(a). A similar behavior is also observed for $MAPbBr_3$ coupled with Ag nanocubes on a Au substrate as shown in Figure 2(b). The difference in the integrated emission intensity of $MAPbBr_3$ coupled with Ag nanocubes on either glass or gold substrates, as depicted in Figure 2(c) indicates that the lasing behavior occurs at a lower threshold (16.5 $W/cm^2$) for the gold substrate than for the glass substrate

(18 W/cm²). This may be due to the gap mode plasmonic resonance, an effect of localized surface plasmon resonance, between Ag nanocubes and underlying gold film [30-32]. In this configuration the Ag nanocubes act as metallic nanostructure while the gold film serves as a substrate and there exists a small gap between metal nanoparticles and gold film. The gap mode plasmonic resonance enhances the scattering and confinement of light within the closed loop formed by multiple scatterings from Ag nanocubes by providing additional feedback paths for photons. This improved feedback mechanism increases the effective path length and enhances the light trapping, resulting in a lower lasing threshold. Additionally, the presence of Au film as a substrate can significantly affect the absorption cross-section of a material as compared to glass substrate due to plasmonic properties of gold [35]. The absorption cross-section is a measure of how efficiently the light is absorbed by the materials at a particular wavelength. Materials having a larger absorption cross-section can also increase the efficiency of population inversion, leading to a lower lasing threshold. Efficient absorption of pump energy is crucial for ensuring that the gain medium has sufficient energy to support the stimulated emission required for lasing.

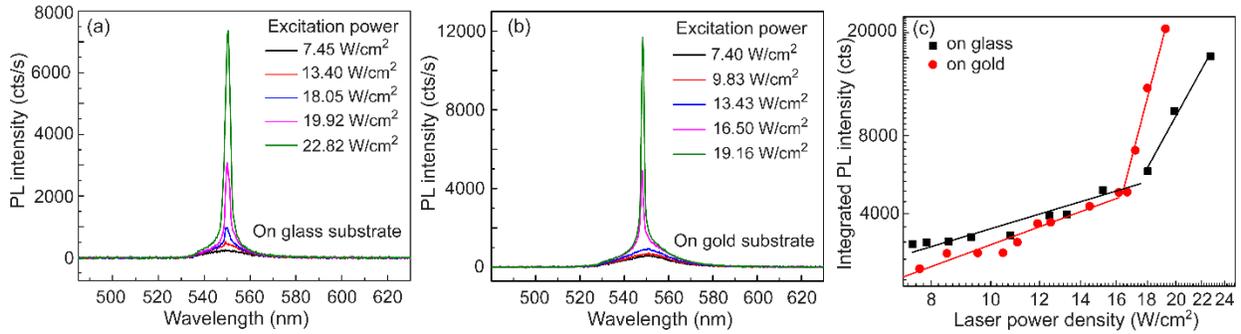

**Figure 2.** PL emission spectra of MAPbBr$_3$/Ag nanocubes under various excitation laser powers at below and above laser threshold on (a) glass substrate and (b) gold substrate. (c) Integrated PL emission intensities of MAPbBr$_3$/Ag nanocubes under various excitation laser powers for both on glass and on gold substrates. Solid lines are linear fits providing guides to the eyes.

In addition, the lasing behavior of MAPbI$_3$ coupled with Ag nanocube on glass substrate was observed as shown in Figure 3(a) and 3(b). Compared to MAPbBr$_3$ when mixed with the same Ag nanocube concentration, the lasing action from MAPbI$_3$ occurs at a lasing threshold much larger (more than double) than that of MAPbBr$_3$. The main reason for this difference is due to the much higher absorption and scattering efficiencies of Ag nanocubes in the visible frequency range [30,32]. It is also evident that for MAPbBr$_3$ the lasing emission peaks near 550 nm (for glass substrate) and 545 nm (for Au substrate) have a narrower FWHM (1 nm) compared to the lasing emission peaks near 782 nm of MAPbI$_3$, which has the FWHM of approximately 5 nm. The difference in the FWHM of the lasing peaks is directly related to the optical confinement, or the quality factor, of the random cavity formed by aggregation of Ag nanocubes as shown in in the inset of Figure 1(a). It is also worth noting that, for both MAPbBr$_3$ and MAPbI$_3$, there are small variations in the lasing peak wavelengths (Figures 2(a,b) and 3(a)). This is particularly different from other random lasing systems such as mixtures of dye/nanostars solutions where both dye molecules and nanoparticles are suspended in liquid [34,36,37]. In our system, the Ag nanocubes are mostly aggregated and adhere to the glass or Au substrate. In other words, the arrangements of Ag nanocube clusters in our study are somewhat fixed, leading to fixed closed loops, and therefore the resonance frequency is unlikely to change during an experiment.

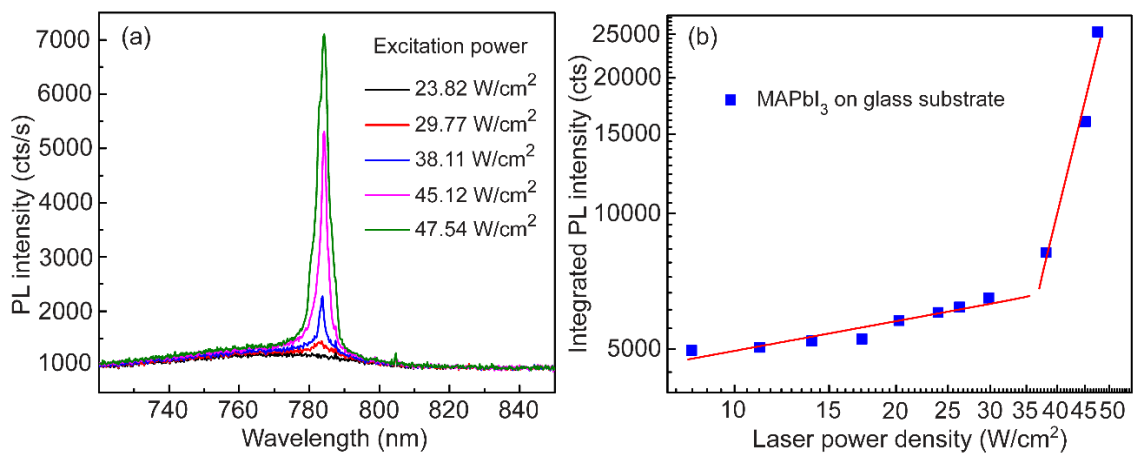

**Figure 3.** (a) PL emission spectra of MAPbI$_3$/Ag nanocubes under various excitation laser powers at below and above laser threshold on a glass substrate. (b) Integrated PL emission intensities of MAPbI$_3$/Ag nanocubes under various excitation laser powers for both on glass substrate. Solid lines are guides to the eyes.

The decay time of optically excited carriers is a critical parameter to distinguish a stimulated emission from a spontaneous emission process. In lasing systems, a shorter decay time is often observed. For MAPbBr$_3$, at the lasing emission wavelength of 550 nm the time resolved measurement of emitted photons showed a shortening of the decay time, from 2.5 ns to 0.45 ns, when the laser power density is increased from below to above the lasing threshold, indicating a transition from spontaneous emission to the stimulated emission regime as shown in Figure 4. It is important to note that due to the limitations of the laser excitation pulse width (~ 0.5 ns), the actual decay time of the lasing emission above the lasing threshold could be much shorter.

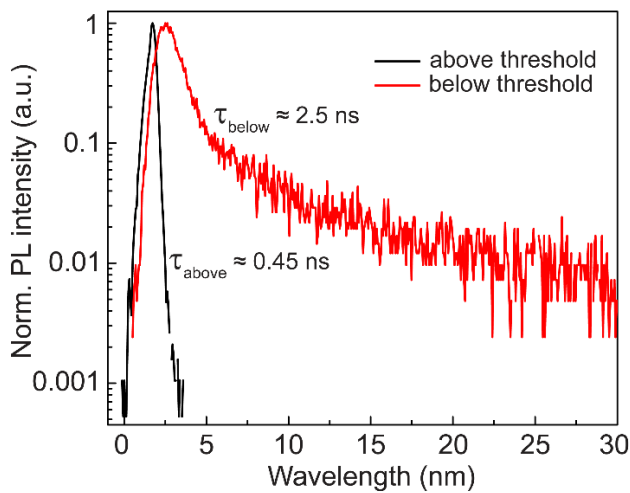

**Figure 4.** Time resolved measurements of MAPbBr$_3$ coupled with Ag nanocube at below and above the lasing threshold.

One of the signatures of a random lasing system with a liquid gain is a fluctuation of the lasing peak intensity over a short time period, even at a fixed excitation fluence above the lasing threshold. This is because of the variation of the material concentration in the laser excitation spot in the solution. At a high laser excitation fluence, perovskite materials can be quenched after approximately 30 seconds to one minute. Subsequently, because of the diffusion of nearby molecules into the laser excitation spot, the lasing behavior continues. Such an observation is shown in Figure 5(a) where the integrated intensity of a lasing peak at a fixed excitation power of 18W/cm is plotted as a function time. Furthermore, it is also noticed that there is an overall decrease in the intensity of the lasing peak over time, which can be understood as a reduction of materials in the vicinity of the excitation spot after a long exposure period. Fluctuations and degradation in the lasing peak intensity present a challenge in determining the lasing efficiency. A potential solution is to embed a mixture of perovskites and Ag nanocubes in a polymer film, which can enhance the stability of the perovskite PL emission.[38]

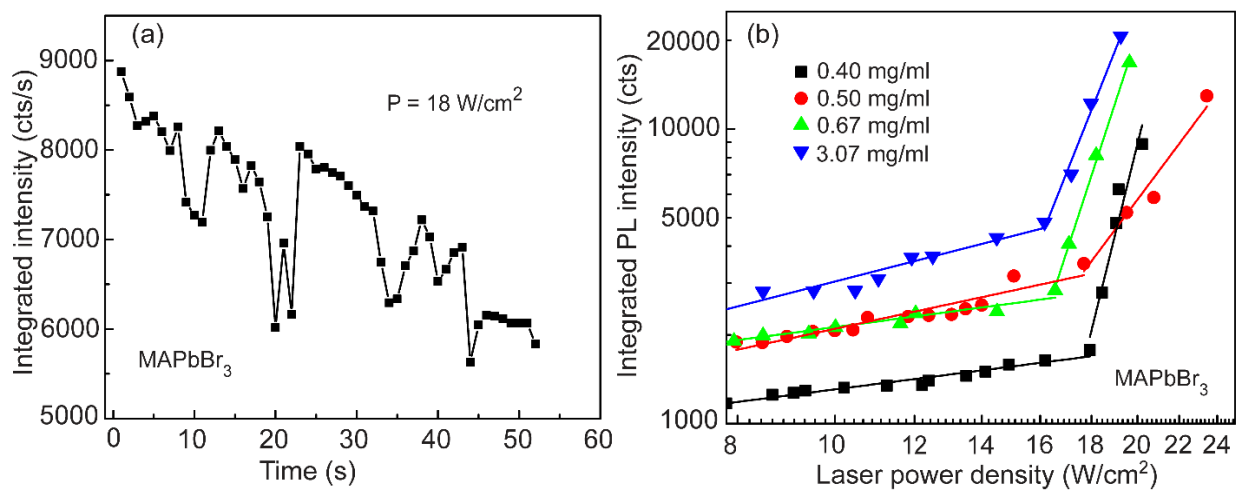

**Figure 5.** (a) Integrated intensity of the lasing emission peak above the lasing threshold as a function of time. (b) Integrated intensity the PL emission under various laser excitation powers for different nanocube concentrations. Solid lines are guides to the eyes.

The power dependence of integrated PL intensity of MAPbBr$_3$ coupled with various concentrations (0.40 mg/ml, 0.50 mg/ml, 0.67 mg/ml, and 3.07 mg/ml) of Ag nanocubes is shown in Figure 5(b) At below the lasing threshold it is observed that the PL intensity is higher for larger concentrations of the Ag nanocubes. Such an enhanced PL spontaneous emission below the lasing threshold is likely due to the plasmon resonances of individual Ag nanocubes. Large spontaneous emission enhancement of fluorophores or inorganic materials by metal nanoparticles through the Purcell effect has been reported previously [39,40]. In Figure 5(b), it also appears that the random lasing threshold is lower for higher concentration of the Ag nanocubes. This can be explained by the fact that for a large concentration of nanocubes the mean free paths of scattered photons become shorter, and then there is a higher probability to form closed loops. Such closed loops can then be amplified, and lasing occurs. Consequently, lasing threshold is influenced by the concentration of scatterer. Previous work by Burin et al. has noted that the lasing threshold is proportional to $N^{-1/2}$ where N scatterer density [41]. It was also observed that the lasing threshold was lowered by more than two orders of magnitude, when the concentration of scattering particles was increased [42].

Finally, we would like to note that one of the primary challenges for random lasing through mirrorless feedback mechanisms is the lack of spatial coherence, and this is also true for the random lasing by perovskites/nanocubes system described here. The absence of well-defined, organized wavefronts can limit the directional control of the emitted light, impacting applications that rely on coherent and controlled light beams [43-45] such as those involving long-distance propagation or controlled directional light emission. Efforts to integrate random lasing with waveguides have shown successful results [10,11,46].

## Summary

In summary, we have demonstrated random lasing action over a broad spectral range by the scattering of emitted photons from $MAPbBr_3$ and $MAPbI_3$ crystalline perovskites with Ag nanocubes. The randomly distributed Ag nanocubes, which act as scatterings centers through their localized surface plasmon resonances, provide strong localized paths for light stimulated amplification. We have successfully demonstrated that the selection of the substrate has an influence on the lasing threshold of $MAPbBr_3$ or $MAPbI_3$ coupled with Ag nanocubes. Additionally, by increasing the concentration of Ag nanocubes in $MAPbBr_3$ or $MAPbI_3$ matrix, we observed lasing at lower laser power densities. The established randomized closed-loop cavities have produced nanometer lasing bandwidths and achieved a low lasing threshold. Furthermore, we have presented a technique for directly steering the random lasing signal into an optical fiber, thus enables a convenient solution to collect and guide the spatially incoherent random lasing through guided modes. The broad plasmonic spectral response of silver nanocubes presented in this work can, in principle, be utilized with different types of gain media.

**Declaration of competing interest**

The authors have no relevant financial or non-financial interests to disclose.

**Data availability**

The data that supports the findings of this study will be made available upon reasonable request.

**Author Contribution**

Yagya Woli: Investigation, Formal analysis, Visualization, Writing – Original draft

Bryson Krause: Investigation, Validation, Resources, Writing – Review and Editing

Thang Hoang: Conceptualization, Validation, Visualization, Resources, Writing – Review and Editing, Supervision.